\documentclass{emulateapj}

\usepackage{natbib}
\usepackage{ulem}
\usepackage{epsfig}
\usepackage{amsmath}

\shorttitle{Metal Outflow in Hydra A}
\shortauthors{Kirkpatrick et al.}

\begin{document}

\title{Direct Evidence for Outflow of Metal-enriched Gas Along the Radio Jets of Hydra A}

\author{C.~C. Kirkpatrick\altaffilmark{1},
              M. Gitti\altaffilmark{2,3,4},
              K.~W. Cavagnolo\altaffilmark{1},
              B.~R. McNamara\altaffilmark{1,4,5},
              L.~P. David\altaffilmark{4},
              P.~E.~J. Nulsen\altaffilmark{4},
              and M.~W. Wise\altaffilmark{6,7}}

\altaffiltext{1}{Department of Physics \& Astronomy, University of Waterloo, 200 University Ave. W., Waterloo, ON N2L 3G1, Canada}
\altaffiltext{2}{INAF - Osservatorio Astronomico di Bologna, via Ranzani 1, I-40127 Bologna, Italy}

\altaffiltext{3}{Department of Astronomy, University of Bologna, via Ranzani 1, I-40127 Bologna, Italy}
\altaffiltext{4}{Harvard-Smithsonian Center for Astrophysics, 60 Garden St., Cambridge, MA 02138}

\altaffiltext{5}{Perimeter Institute for Theoretical Physics, 31 Caroline St. N., Waterloo, ON N2L 2Y5, Canada}
\altaffiltext{6}{ASTRON, Netherlands Institute for Radio Astronomy, Postbus 2, 7990AA Dwingeloo, The Netherlands}
\altaffiltext{7}{Astronomical Institute ``Anton Pannekoek", University of Amsterdam, P.O. Box 94249, 1090GE Amsterdam, The Netherlands}

\begin{abstract}

Using deep {\it Chandra} observations of the Hydra A galaxy cluster, we examine the metallicity structure near the central galaxy and along its powerful radio source.  We show that the metallicity of the intracluster medium is enhanced by up to 0.2 dex along the radio jets and lobes compared to the metallicity of the undisturbed gas.  The enhancements extend from a radius of 20 kpc from the central galaxy to a distance of $\sim 120$ kpc.  We estimate the total iron mass that has been transported out of the central galaxy to be between 2$\times$10$^7$ M$_\sun$ and 7$\times$10$^7$ M$_\sun$ which represents 10\% - 30\% of the iron mass within the central galaxy.   The energy required to lift this gas is roughly 1\% to 5\% of the total energetic output of the AGN.  Evidently, Hydra A's powerful radio source is able to redistribute metal-enriched, low entropy gas throughout the core of the galaxy cluster. The short re-enrichment timescale $< 10^9$ yr implies that the metals lost from the central galaxy will be quickly replenished.

\end{abstract}


\section{Introduction}

The elemental abundances of the hot gas at the centers of cooling flow clusters are often enhanced with respect to those at larger radii \citep{deg01}.  The high central iron abundance relative to the $\alpha$-elements suggests the enhancements are largely composed of ejecta from type Ia supernova explosions associated with brightest cluster galaxies (BCG) \citep{deg04,tam04}.  Radial metallicity profiles of the hot gas tend to be broader than the stellar light profiles of BCGs, indicating that metal-enriched gas is diffusing outward and mixing with the intracluster medium \citep{reb05,reb06,dav08,ras08}.  BCGs frequently harbor active galactic nuclei (AGN) which are able to heat the metal enriched gas surrounding them, causing it to expand and mix with lower metallicity gas at larger radii \citep{dav08}.  In addition, radio sources associated with AGN may be able to entrain metal enriched gas and transport it outward anisotropically along the radio jets and lobes \citep{gop01,gop03,bru02,omm04,hea07,roe07}.  

Inclusions of unusually high metallicity gas have been found outside the centers of a handful of groups and clusters \citep{san04,gu07,sim08}, but their association with AGN has not been firmly established.  Evidence linking a metal enriched outflow to jet activity has been found in the Hydra A cluster which shows enhanced metallicity near its radio lobes in an {\it XMM-Newton}  X-ray image \citep{sim09}.  Hydra A has an extensive system of cavities embedded in its ICM that is filled with radio emission emanating from the nucleus of the BCG \citep{bmc00,dav01,nul02,nul05,wis07}.  With a total energy of over 10$^{61}$ erg, Hydra A is experiencing one of the most powerful AGN outbursts in the nearby universe.  It is thus an ideal candidate to search for and characterize a metal-enriched outflow.

We performed such an analysis on a deep, high-resolution X-ray image obtained with the {\it Chandra X-ray Observatory}, and we find clear-cut evidence for metal enriched gas aligned  with the radio source.  Throughout this letter we assume a $\Lambda$CDM cosmology with H$_{0} =$ 70 km s$^{-1}$ Mpc$^{-1}$, $\Omega_{M} =$ 0.3, and $\Omega_{\Lambda} =$ 0.7.  Hydra A is at a redshift of $z= 0.055$, which corresponds to an angular scale of 1.07 kpc arcsec$^{-1}$.  All uncertainties are quoted at the 90\% confidence level.

\begin{figure*}[t]
\begin{center}
$\begin{array}{ccc}
\leavevmode \epsfysize=6cm \epsfbox {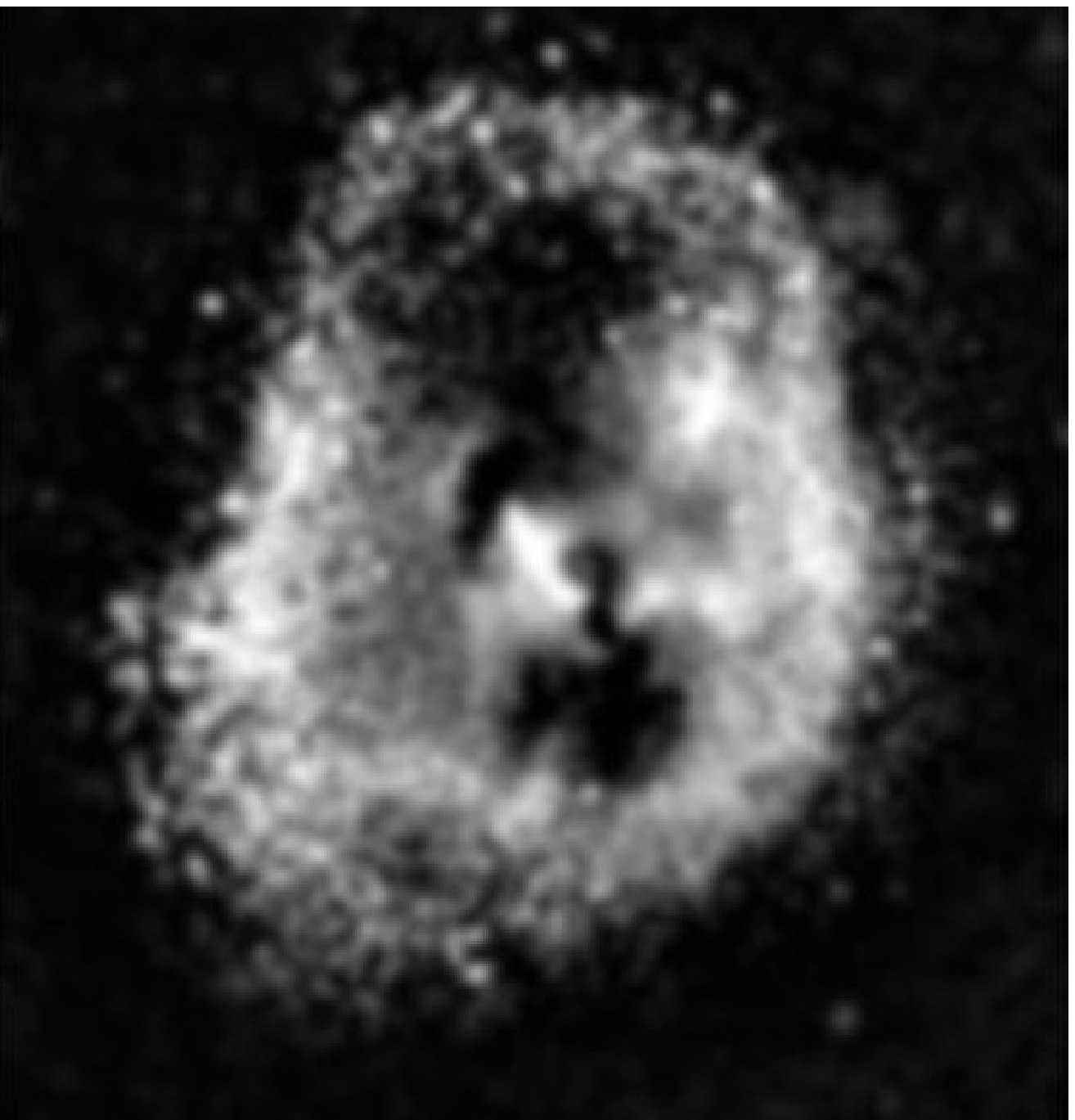} &
\leavevmode \epsfysize=6cm \epsfbox {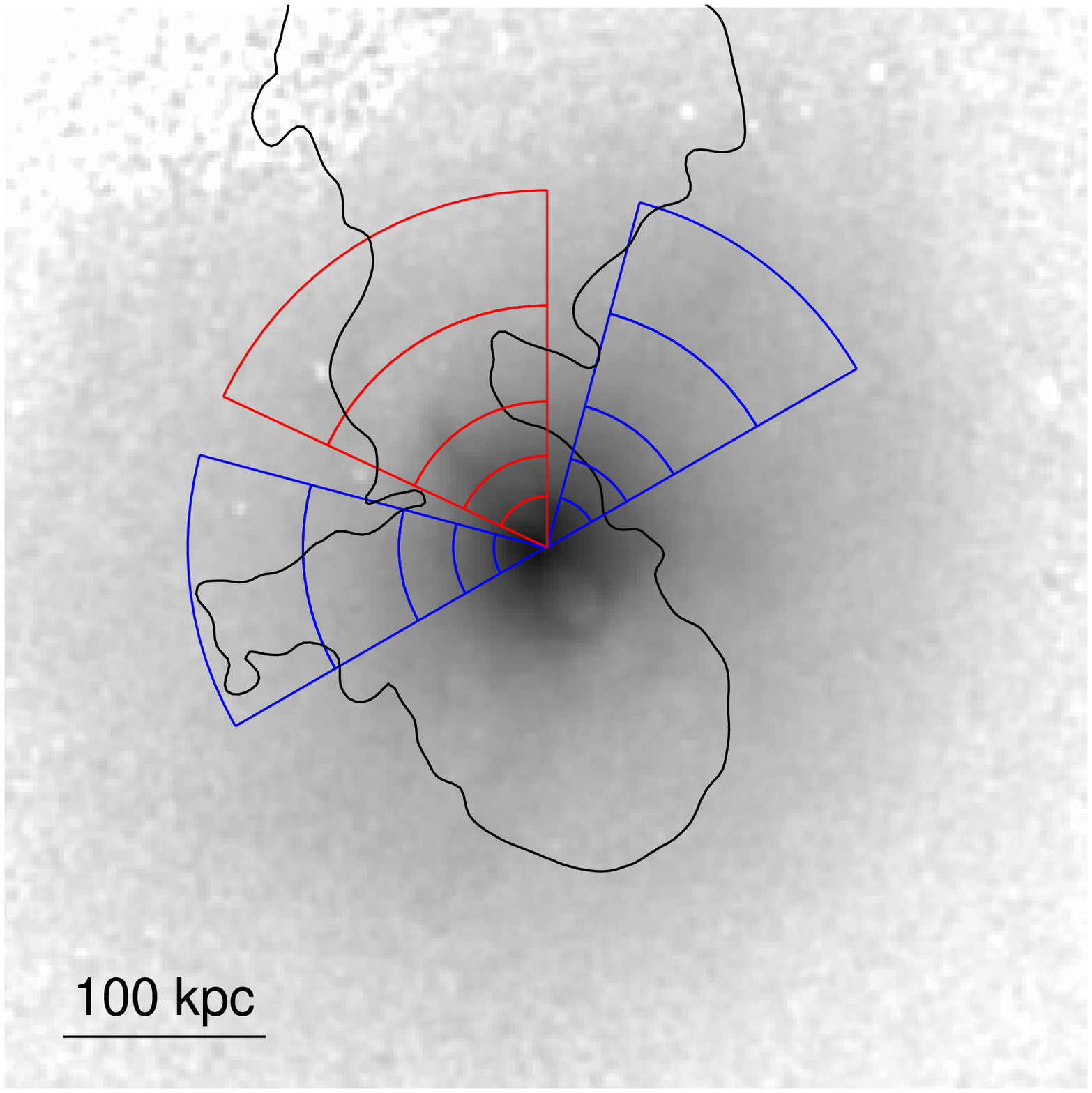} &
\leavevmode \epsfysize=6cm \epsfbox {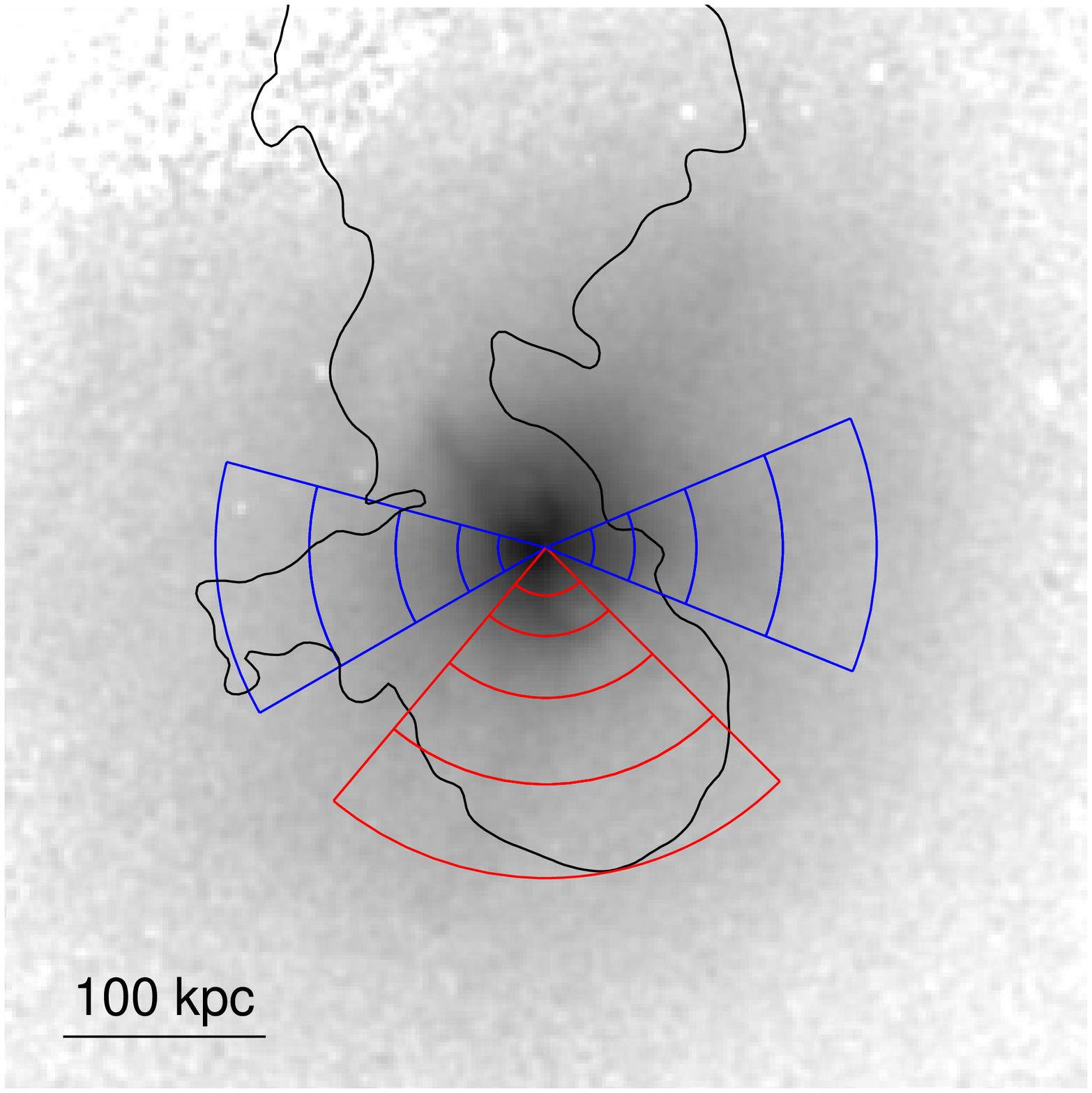}\\
\end{array}
$
\caption{First panel: Residual map of the beta model subtracted surface brightness image from \citet{wis07}.  North is towards the top, east is towards the left.  Second panel: Combined, adaptively smoothed image of Hydra A.  The black contour outlines the 330 MHz radio emission.  The red regions are where the spectra were extracted along the northern jet and the blue regions are where the off-axis spectra was extracted.  Each bin contains at least 44000 counts.  Third panel:  Same as the second panel, but for the southern jet.}
\end{center}
\end{figure*}

\section{Data Analysis}

\subsection{X-ray Data Reduction}

{\it Chandra} X-ray observations 4969 and 4970 were processed using CIAO version 4.0.1 and version 3.4.5 of the calibration database.  Background flares were excluded using standard filtering, yielding a net exposure time of 182 ks.  Time-dependent gain and charge transfer inefficiency corrections were applied.  Blank-sky background files used for background subtraction were normalized to the source count rate in the 9.5 - 12 keV band.

\subsection{Metallicity Profile Along the Jets}

We have searched for a metal-enriched outflow by comparing the radial distribution of iron along and orthogonal to the radio jets and lobes.  Guided by the cavity system and 330 MHz radio emission shown in Figure 1, we have defined sectors parallel to and roughly orthogonal to the jets.  The northern sector located between 90 and 155 degrees, measured counterclockwise from west, was divided into five radial bins with minimum signal-to-noise of 210.  Two regions east and west of the northern jet were created with 45 degree opening angles and bin sizes matching those of the northern sector.  These regions are outlined in Figure 1.  The spectrum for each bin was extracted and corresponding event-weighted response matrices were computed.  Using XSPEC \citep{arn96}, single temperature plasma models with absorption (WABS$\times$MEKAL) were fit to the spectra in the energy range 0.5 to 7 keV.  The column density was frozen at the Galactic value of $4.94 \times 10^{20}$ cm$^{-2}$ \citep{dic90}.  Temperatures, abundances, and normalizations were allowed to vary, while the abundance ratios were set to the the solar photospheric values of \citet{gre98}.  The spectra are very well fit by this model.

The projected metallicity profiles along and orthogonal to the northern jet are shown in the upper panel of Figure 2.  Between radii of $\sim 20$ kpc and 120 kpc  the metal abundance of the hot gas is enhanced by up to 0.2 dex along the jet relative to the orthogonal sectors.  Metallicities for the inner and outermost bins are indistinguishable between the on and off jet sectors.  

The southern sector is located between 230 and 315 degrees counterclockwise from west.  The metallicity profiles along and orthogonal to the southern jet are shown in the lower panel of Figure 2.  A metallicity enhancement is also found to the south, extending to approximately 120 kpc.  The lower points in Figure 2 shown as triangles represent  the orthogonal sectors and are interpreted as the undisturbed metallicity gradient.  The gas in the undisturbed regions is warmer by up to 0.5 keV compared to the jet regions.

The metallicity enhancement could be affected by the presence of cooler gas.  To explore this, we have considered a second temperature component when fitting the regions along the jets.  For example, a component with a temperature between 0.5 and 0.7 keV in each sector would increase the abundances by approximately 15\%, but would contribute at most 2\% to the total emission.  However, this and other similar models do not significantly improve the spectral fit.  If such an extended cool component were present, it would have the effect of increasing the metal enhancements shown in Figure 2 by roughly 0.1 dex.

\begin{figure}[b]
\epsscale{1.25}
\plotone{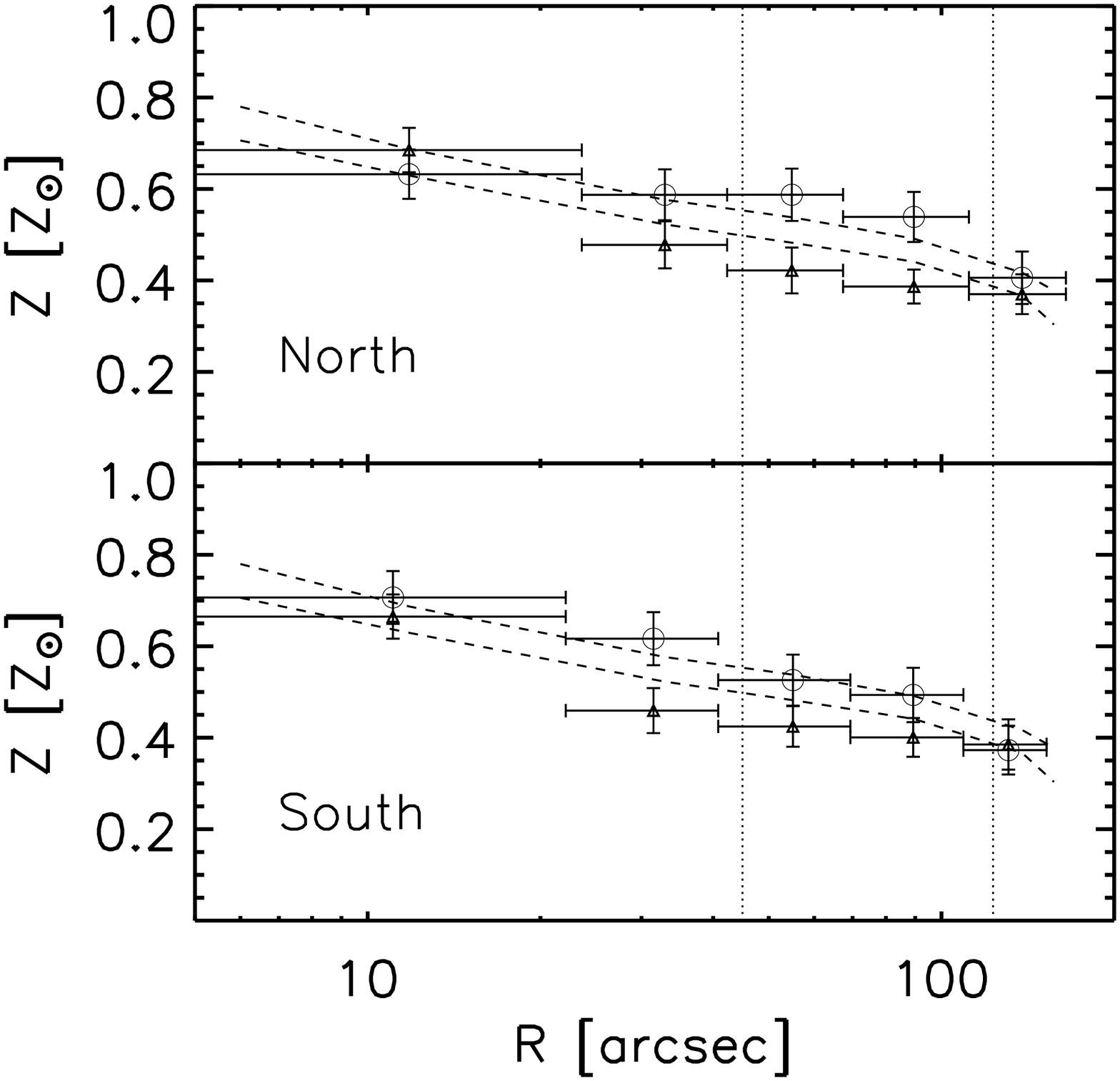}
\caption{Projected metallicity profiles for Hydra A.  The circles represent regions along the jets and triangles are the average fits to the regions east and west of each jet.  The dashed lines represent the aximuthally-averaged metallicity profile.  The left vertical dotted lines indicate the outer edges of the cavity pairs A and B from \citet{wis07}, and the right dotted line is for cavity pair C and D.}
\end{figure}

\begin{figure}
\epsscale{1.15}
\plotone{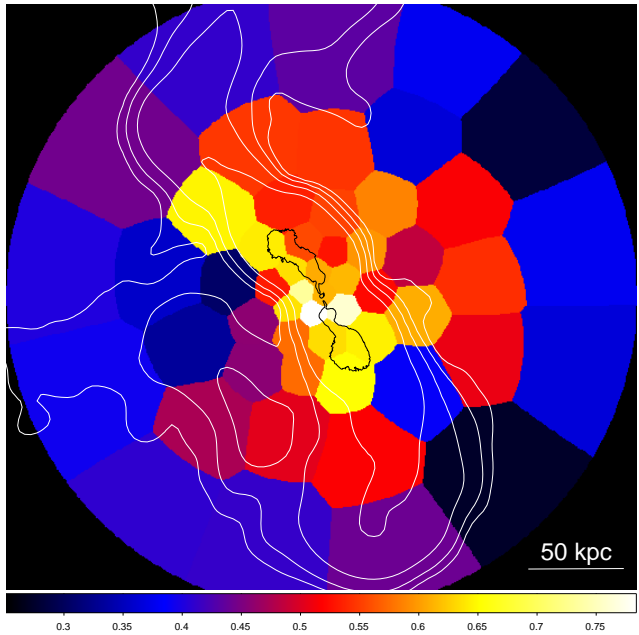}
\caption{Metallicity map showing the central 5' $\times$ 5' of Hydra A.  Each bin contains approximately 22500 counts.  Brighter regions represent a higher metallicity.  The average error per bin is approximately 18\%.  The 330 MHz emission is shown by the white contours and the 1400 MHz emission is shown by the black contours.}
\end{figure}

\subsection{Metallicity Map}

The abundance map shown in Figure 3 was created using a weighted Voronoi tessellation binning algorithm \citep{cap03,die06}.  Each bin has a S/N of 150  and was chosen to balance the uncertainties on the best-fit abundances while preserving the excellent spatial resolution afforded by the large number of source counts.  For each bin, a spectrum was extracted using the procedure described in Section 2.2.  The spectrum from each bin was fitted in the energy range 0.5-7.0 keV with an absorbed single temperature MEKAL thermal plasma model using the parameters defined previously in Section 2.2.

The highest metallicity gas, shown in yellow in Figure 3, extends to the north-east and south of the nucleus.  This gas is closely aligned with the radio emission and is consistent with the metallicity excesses shown in Figure 2.  The temperature along the north-east jet is approximately 2.7 keV, and the southern jet is found to be approximately 3.1 keV.  The surrounding region has a temperature that ranges between 3.4 keV and 3.8 keV.  This is similar to what was found in the {\it XMM} analysis by \citet{sim09}.  A plume of uplifted gas with a modest metallicity enhancement of $\sim 0.1$ Z/Z$_\sun$, shown in orange, can be seen extending north-west along an X-ray filament. Though it is located within what we have defined as the undisturbed region described in section 2.2, it has a negligible effect on our analysis.

\section{Discussion}

\subsection{Iron Mass}

We estimate the excess iron mass in two regions.  First, we assume the metal-enriched gas fills the entirety of the on-jet regions defined in Figure 1.  Second, we consider only in the most significant iron excess shown in yellow in Figure 3.  We consider both regions because the mass estimate depends critically on the volumes and filling factors (which we assume to be unity) of the metal-enhanced regions.  The uncertainty associated with these quantities should be smaller in the yellow region alone.  At the same time, the true enhancement must be larger as it is indeed spread over a large volume.  We compute the iron mass as
\begin{equation}
M_{\rm Fe} = \rho V Z f_{{\rm Fe},\sun}, 
\end{equation}
where $Z$ is the metallicity of the gas, $f_{{\rm Fe},\sun}$ is the iron mass fraction of the Sun, $V$ is the volume, and $\rho$ is the density of the gas assuming $n_{\rm e} = 1.2 n_{\rm H}$.  The density was calculated from the deprojection of the surface brightness profile along the jets assuming the volumes of the semi-annular wedges shown in Figure 1.  

\begin{figure}[t]
\epsscale{1.25}
\plotone{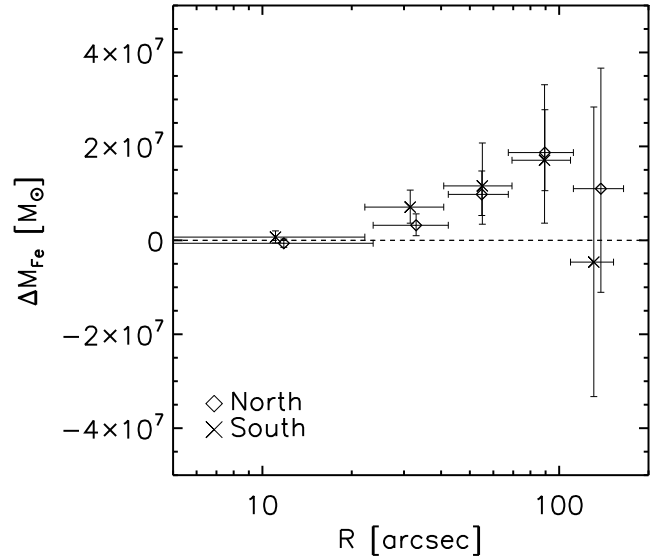}
\caption{The diamond points represent the excess iron mass along the northern jet compared to the global profile.  The cross points represent the excess iron mass along the southern jet.  Only the second, third, and fourth bins show a significant excess.}
\end{figure}

For the first case the difference between the iron mass along the jet and the iron mass of the underlying metal distribution in equal volumes is plotted in Figure 4 for both north and south regions.  A significant excess is seen in the second, third, and fourth bins along both jets.  The total volume of these bins combined is $4.1 \times 10^{70}$ cm$^3$.  We find a total excess iron mass of $3.2_{-1.0}^{+1.1} \times 10^7$ M$_{\sun}$ along the northern jet and $3.6_{-1.6}^{+1.9} \times 10^7$ M$_{\sun}$ along the southern jet.  These values are consistent with the upper limit found by \citet{sim09}.  

In our second case, the volumes of the regions are approximated as cylinders extending from 15 kpc outside the nucleus to a radius of 100 kpc, with an approximate width of 25 kpc.  Their total volume is $2.7 \times 10^{69}$ cm$^3$.  After subtracting the average metallicity of the undisturbed background 0.42 Z/Z$_\sun$, we estimate the excess iron mass within the yellow region in Figure 3 to be $\sim 1.6 \times 10^7$ M$_\sun$.  

The excesses iron mass represents a significant fraction of the total iron mass in the gas surrounding the BCG.  The total iron mass measured within a 30 kpc radius is $1.61_{-0.52}^{+0.60} \times 10^8$ M$_\sun$, consistent with the value in both \citet{dav01} and \citet{sim09}.  Assuming the the excess iron originated at the center of the cluster, this implies that $\sim$ 10\% - 30\% of the iron originating from this central region has been displaced along the jets.

Using the relationship of \citet{boh04}, we estimate the time required to replenish the lost iron through SNe Ia and stellar mass loss as,
\begin{equation}
t_{\rm enr} =  \left ( 10^{-12} S \eta_{\rm Fe} + 2.5\times10^{-11} \gamma_{\rm Fe}\right ) ^{-1} \frac{M_{\rm Fe} L_{{\rm B}\sun}}{L_{\rm B}}.
\end{equation}
Here, $\eta_{\rm Fe} = 0.7$ M$_\sun$ is the iron yield from SN Ia and $\gamma_{\rm Fe} = 2.8 \times 10^{-3}$ is the iron mass fraction from stellar mass loss.  With a $B$-band luminosity of L$_{\rm B} = 9.2 \times 10^{10}$ L$_\sun$ \citep{dav01} and a supernova rate of $S = 0.15$ SNU \citep{cap99}, the enrichment time required to replace the uplifted iron is approximately 0.2 - 0.7 Gyr.  The enrichment time is a small fraction of the age of the cluster and is only two to seven times larger than the age of the outburst that created the surrounding shock front.  Thus, the central metallicity peak would recover if the AGN becomes dormant for a period of only a few times its current age.

Because some star formation is occurring in the BCG, we have also considered the contribution from SN II enrichment.  Assuming one supernova per 100 solar masses of stars produced and a star formation rate of 1 - 5 M$_\sun$ yr$^{-1}$ \citep{bmc95} in the BCG, we find only 3\% of the total iron mass is expected to originate from SN II enrichment over this time-scale.  




\subsection{Outflow Energy}

The fraction of the AGN outburst energy required to lift the enriched gas to its present position provides a lower limit to the AGN energy deposited into the high metallicity gas near the BCG.  This value can be estimated by calculating the difference in gravitational potential energy between the original position of the uplifted gas and its current position.  Following \citet{rey08}, we calculate this quantity as,
\begin{equation}
\Delta E = \frac{M c_{\rm s}^2}{\gamma} \ln \left( \frac{\rho_i}{\rho_f} \right),
\end{equation}
where the sound speed is c$_{\rm s} \approx 750$ km s$^{-1}$, $\rho_i$ and $\rho_f$ are the initial and final densities of the surrounding ICM and $\gamma = 5/3$ is the ratio of specific heat capacities.  Assuming all of the displaced gas was initially at the center of the cluster, the energy required to lift the metal-enriched gas lies between 0.5 - 5$\times$10$^{59}$ erg.  This figure is comparable to the work required to inflate the inner cavities against the surrounding pressure \citep{wis07} and is between 1\% and 5\% of the total energy expended by all of the cavities and the surrounding shock front.  This analysis suggests that lower power AGN outbursts typically observed in clusters \citep{bmc07} would have a minor impact on their metallicity distributions.

\section{Conclusions}

Hydra A has undergone a major AGN outburst that has transported 2 - 7$\times$10$^7$ M$_\sun$ of metal-enriched gas out of the BCG along the radio jets.  The gas has been enhanced by up to 0.2 dex in metallicity compared to the ambient gas out to distances as large as 120 kpc.  Approximately 10$^{59}$ erg was expended while uplifting the metal-enriched gas, which is 1\% to 5\% of the total AGN outburst energy.  

\acknowledgments

This work was supported by generous grants from the Natural Sciences and Engineering Research Council of Canada and {\it Chandra} grant G07-8122X.  M. Gitti acknowledges support by grant ASI-INAF I/088/06/0.


\end{document}